\documentclass[natbib,preprint]{sigplanconf}
\usepackage{alltt}
\usepackage{subfigure}
\usepackage{amsmath}
\usepackage{amssymb}
\usepackage{amsthm}
\usepackage{array}
\usepackage[pdftex]{graphicx}
\usepackage{semantic}
\usepackage{multirow}
\usepackage{paralist}
\usepackage{stmaryrd}
\usepackage{url}
\usepackage[all]{xy}
\usepackage{todonotes}

\newcommand{\rdot}{{\boldsymbol{.}}}

\newcommand{\htree}[3]{#1 \to_{#3} #2}

\newcommand{\G}{\ensuremath{\mathcal{G}}}
\newcommand{\C}{\ensuremath{\mathcal{C}}}

\newcommand{\ptrees}[0]{\mathcal{T}_{\mathcal{G}}}
\newcommand{\filter}[1]{\mathcal{F}(#1)}

\theoremstyle{definition}
\newtheorem{definition}{Definition}[section]
\newtheorem{example}[definition]{Example}

\title{Well-typed Islands Parse Faster}

\authorinfo{Erik Silkensen}
           {University of Colorado at Boulder}
           {erik.silkensen@colorado.edu}
\authorinfo{Jeremy Siek}
           {University of Colorado at Boulder}
           {jeremy.siek@colorado.edu}

\begin{document}
\maketitle

\begin{abstract}
  This paper addresses the problem of specifying and parsing the
  syntax of domain-specific languages (DSLs) in a modular,
  user-friendly way.  That is, we want to enable the design of
  \emph{composable} DSLs that combine the natural syntax of external
  DSLs with the easy implementation of internal DSLs.
  The challenge in parsing composable DSLs is that the composition
  of several (individually unambiguous) languages is likely to contain
  ambiguities.
  In this paper, we present the design of a system that uses a
  type-oriented variant of island parsing to efficiently parse the
  syntax of composable DSLs.
  In particular, we show how type-oriented island parsing is
  constant time with respect to the number of DSLs imported. We also
  show how to use our tool to implement DSLs on top of a host language
  such as Typed Racket.
\end{abstract}

\section{Introduction}

Domain-specific languages (DSLs) provide high productivity for
programmers in many domains, such as computer systems, physics, linear
algebra, and other sciences. However, a series of trade-offs face the
prospective DSL designer today. On the one hand, external DSLs offer
natural syntax and friendly diagnostics at the cost of
interoperability issues~\cite{Beazley1996} and difficulty of
implementation.  They are usually either implemented by hand or by
using parser generators \emph{\`a la} {\sc yacc} that require technical
knowledge of parsing algorithms.  Meanwhile, many general-purpose
languages include a host of tricks for implementing internal (or
embedded) DSLs, such as templates in C++, macros in Scheme, and type
classes in Haskell; however, the resulting DSLs are often \emph{leaky
  abstractions}: the syntax is not quite right, compilation errors
expose the internals of the DSL, and debuggers are not aware of the
DSL.

In this paper, we make progress towards combining the best of both
worlds into what we call \emph{composable} DSLs. We want to enable
fine-grained mixing of languages with the natural syntax of external
DSLs and the interoperability of internal DSLs.

\begin{figure}
  \begin{center}
    \includegraphics[scale=0.5]{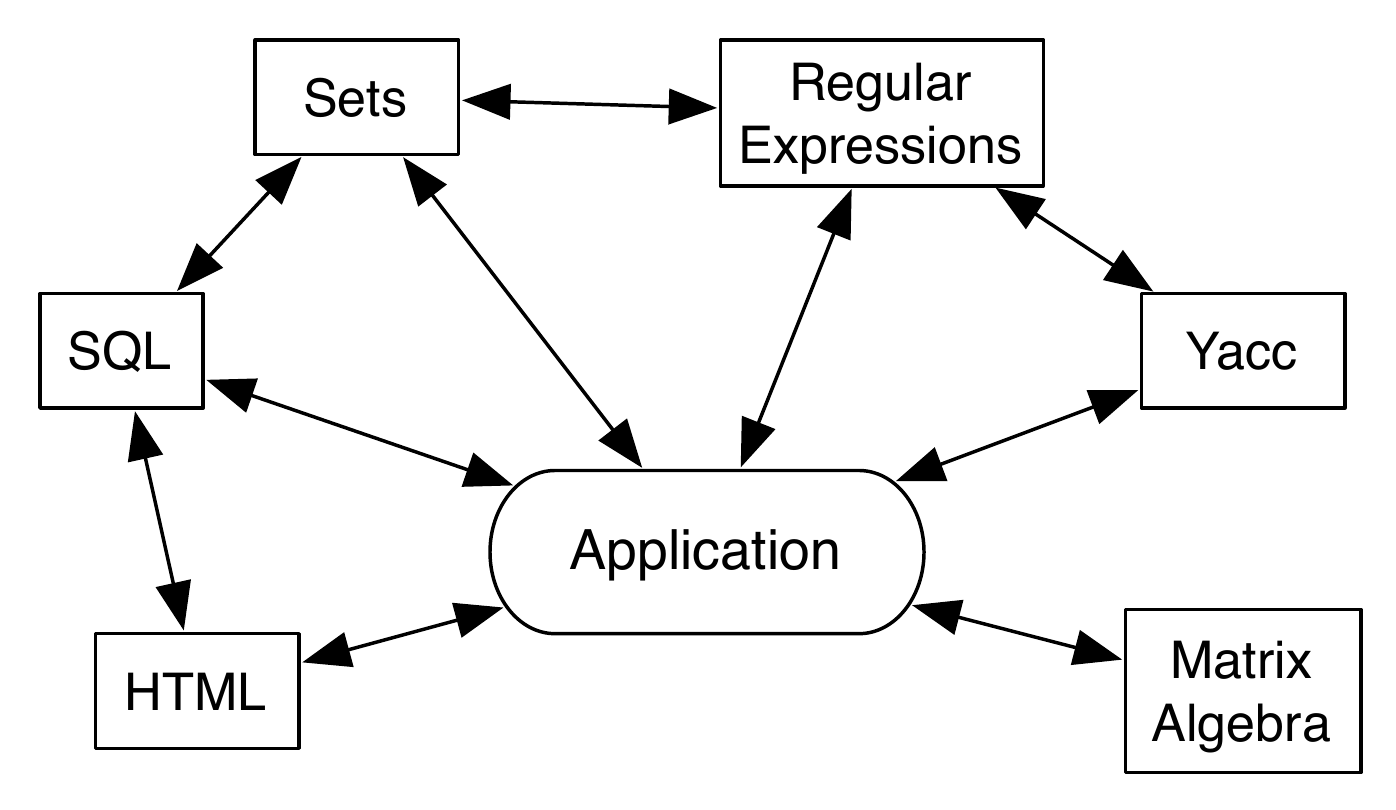}
  \end{center}
  \caption{Our common case: an application using many DSLs.}
  \label{composing}
\end{figure}

At the core of this effort is a parsing problem: although the grammar
for each DSL may be unambiguous, programs that use multiple DSLs,
such as the one in Figure~\ref{composing}, need to be parsed using the
union of their grammars, which are likely to contain
ambiguities~\cite{Kats:2010}.  Instead of relying on the grammar
author to resolve them (as in the LALR tradition), the parser
for such an application must be able to efficiently deal with
ambiguities.


We should emphasize that our goal is to create a parsing system that
provides much more syntactic flexibility than is currently offered
through operator overloading in languages such as C++ and Haskell.  We
are not trying to build a general purpose parser, that is, we are
willing to place restrictions on the allowable grammars, so long as
those restrictions are easy to understand (for our users) and do not
interfere with composability.

As a concrete, motivating example, we consider the union of grammars
for matrix algebra, regular expressions, and sets outlined in
Figure~\ref{composing-amb}.  Written in the traditional style, the
union of these individually unambiguous grammars is greatly ambiguous;
so importing many DSLs such as these can increase the parse time by
orders of magnitude even though the program is otherwise unchanged.
Of course, an experienced computer scientist will immediately say that
the separate grammars should be merged into one grammar with only one
production for each operator. However, that would require coordination
between the DSL authors and is therefore not scalable.

\begin{figure}
\begin{alltt}
{\bf{module}} MatrixAlgebra \{
  Expr ::= Expr "+" Expr [left,1]
         | Expr "-" Expr [left,1]
         | Expr "*" Expr [left,2]
         | "|" Expr "|" | Id; \(\cdots\)
\}
{\bf{module}} RegularExpressions \{
  Expr ::= "'" Char "'" | Expr "+" | Expr "*" 
         | Expr "|" Expr [left] | Id; \(\cdots\)
\}
{\bf{module}} Sets \{
  Expr ::= Expr "+" Expr [left,1]
         | Expr "-" Expr [left,2] | Id; \(\cdots\)
\}

{\bf{import}} MatrixAlgebra, RegularExpressions, Sets;
A + B + C     {\it{// Ambiguous!}}
\end{alltt}
  \caption{Ambiguity due to the union of DSLs.}
  \label{composing-amb}
\end{figure}

\subsection{Type-Oriented Grammars}

To address the problem of parsing composed DSLs, we observe that
different DSLs define different types: \emph{Matrix}, \emph{Vector},
and \emph{Scalar} in Matrix Algebra, \emph{Regexp} in Regular
Expressions, \emph{Set} in Sets, and so on.
We suggest an alternate style of grammar organization that we call
\emph{type-oriented grammars}, inspired by \citet{Sandberg1982}. In
this style, a DSL author creates one nonterminal for each type in the
DSL and uses the most specific nonterminal/type for each operand in a
grammar rule.
Figure~\ref{fig:composing-disamb} shows the example from
Figure~\ref{composing-amb} rewritten in a type-oriented style, with
nonterminals for \texttt{Matrix}, \texttt{Vector}, \texttt{Scalar},
\texttt{Regexp}, and \texttt{Set}.

\subsection{Type-based Disambiguation}

While the union of the DSLs in Figure~\ref{fig:composing-disamb} is no
longer itself ambiguous, programs such as $\mathtt{A + B + C} \cdots$
are still highly ambiguous if the variables \texttt{A}, \texttt{B},
and \texttt{C} can each be parsed as either \texttt{Matrix},
\texttt{Regexp}, or \texttt{Set}.  Many prior
systems~\cite{Paulson:1994vn,Bravenboer2005b} use chart
parsing~\cite{Kay1986} or GLR~\cite{Tomita1985} to produce a parse
forest and then type check to filter out the ill-typed
trees. This solves the ambiguity problem, but these parsers are
inefficient on ambiguous grammars (Section~\ref{sec:evaluation}).

This is where our key contribution comes in: \emph{island parsing with
  eager, type-based disambiguation}. We use a chart parsing strategy,
called island parsing~\cite{Stock1988} (or bidirectional bottom-up
parsing~\cite{Quesada:1998zr}), that enables our algorithm to grow
parse trees outwards from \emph{well-typed terminals}. The statement
\begin{center}
  \textbf{declare} \texttt{A:Matrix, B:Matrix, C:Matrix} \{ \ldots \}
\end{center}
gives the variables \texttt{A}, \texttt{B}, and \texttt{C} the type
\texttt{Matrix}. We then integrate type checking into the parsing
process to prune ill-typed parse trees before they have a chance to
grow, drawing inspiration from from the field of natural language
processing, where using types to resolve ambiguity is known as
\emph{selection restriction}~\cite{Jurafsky2009},

Our approach does not altogether prohibit grammar ambiguities; it
strives to remove ambiguities from the common case when composing DSLs
so as to enable efficient parsing.

\begin{figure}
\begin{alltt}
{\bf{module}} MatrixAlgebra \{
  Matrix ::= Matrix "+" Matrix [left,1]
           | Matrix "-" Matrix [left,1]
           | Matrix "*" Matrix [left,2];
  Scalar ::= "|" Vector "|"; \(\cdots\)
\}
{\bf{module}} RegularExpressions \{
  Regexp ::= "'" Char "'" | Regexp "+" 
          | Regexp "*" | Regexp "|" Regexp; \(\cdots\)
\}
{\bf{module}} Sets \{
  Set ::= Set "+" Set [left,1]
        | Set "-" Set [left,2]; \(\cdots\)
\}

{\bf{import}} MatrixAlgebra, RegularExpressions, Sets;
{\bf{declare}} A:Matrix, B:Matrix, C:Matrix \{ 
  A + B + C
\}
\end{alltt}
  \caption{Type-oriented grammars for DSLs.}
  \label{fig:composing-disamb}
\end{figure}



\subsection{Contributions}

\begin{enumerate}
\item We present the first parsing algorithm, \emph{type-oriented
    island parsing} (Section~\ref{sec:type-island}), whose time
  complexity is \emph{constant} with respect to the number of DSLs in
  use, so long as the nonterminals of each DSL are largely disjoint
  (Section~\ref{sec:evaluation}).

\item We present our extensible parsing system\footnote{
   See the supplemental material for the URL for the code.
  } that adds several features to the parsing algorithm to make it
  convenient to develop DSLs on top of a host language such as Typed
  Racket~\cite{Tobin-Hochstadt2008}
  (Section~\ref{sec:bells-whistles}).  

  
\item We demonstrate the utility of our parsing system with an example
  in which we embed syntax for two DSLs in Typed Racket.
\end{enumerate}

Section~\ref{sec:background} introduces the basic definitions and
notation used in the rest of the paper.
We discuss our contributions in relation to the prior literature in
Section~\ref{sec:related-work} and conclude in
Section~\ref{sec:conclusions}.


\section{Background}
\label{sec:background}

We review the definition of a grammar and parse tree and present our
framework for comparing parsing algorithms, which is based on the
parsing schemata of \citet{Sikkel1998}.

\subsection{Grammars and Parse Trees}

A \emph{context-free grammar} (CFG) is a 4-tuple $\mathcal{G} = (\Sigma,
\Delta, \mathcal{P}, S)$ where $\Sigma$ is a finite set of terminals, $\Delta$ is
a finite set of nonterminals, $\mathcal{P}$ is finite set of grammar
rules, and $S$ is the start symbol. We use $a, b, c,$ and $d$ to range
over terminals and $A, B, C,$ and $D$ to range over nonterminals.
The variables $X,Y,Z$ range over symbols, that is, terminals and nonterminals,
and $\alpha, \beta, \gamma, \delta$ range over sequences of symbols. Grammar rules
have the form $A \to \alpha$. We write $\G{} \cup (A \to \alpha)$ as an abbreviation
for $(\Sigma,\Delta,\mathcal{P} \cup (A \to \alpha),S)$.

We are ultimately interested in parsing programs, that is, converting
token sequences into abstract syntax trees. So we are less concerned
with the recognition problem and more concerned with determining the
parse trees for a given grammar and token sequence.
The \emph{parse trees} for a grammar $\G = (\Sigma,\Delta,\mathcal{P},S)$,
written $\mathcal{T}(\G)$, are trees built according to the following
rules.
\begin{enumerate}
\item If $a \in \Sigma$, then $a$ is a parse tree labeled with $a$.
\item If $t_1,\ldots,t_n$ are parse trees labeled $X_1,\ldots,X_n$
  respectively, $A \in \Delta$, and $A \to X_1,\ldots,X_n \in \mathcal{P}$, then
  the following is a parse tree labeled with $A$.
\[\footnotesize
\xymatrix{
     & A \ar@{-}[dl] \ar@{-}[dr] &  \\
  t_1 & \cdots & t_n
}
\]
\end{enumerate}
We sometimes use a horizontal notation $A \to t_1\ldots t_n$ for parse
trees and we often subscript parse trees with their labels, so $t_A$
is parse tree $t$ whose root is labeled with $A$. We use an overline
to represent a sequence: $\overline{t} = t_1,\ldots,t_n$.

The \emph{yield} of a parse tree is the concatenation of the labels on
its leaves:
\begin{align*}
  \mathit{yield}(a) &= a \\
  \mathit{yield}([A \to t_1\ldots t_n]) &=  \mathit{yield}(t_1) \ldots \mathit{yield}(t_n)
\end{align*}

\begin{definition}
  The set of parse trees for a CFG $\G{} = (\Sigma,\Delta,\mathcal{P},S)$ and
  input $w$, written $\mathcal{T}(\G,w)$, is defined as follows
\[
  \mathcal{T}(\G,w) = \{ t_S \mid t_S \in \mathcal{T}(\G) \text{ and } 
      \mathit{yield}(t_S) = w \} 
\]
\end{definition}

\begin{definition}
The \emph{language} of a CFG \G{}, written $L(\mathcal{G})$, consists
of all the strings for which there is exactly one parse tree. More
formally,
\[
L(\G) = \{ w \mid | \mathcal{T}(\G,w)| = 1 \} 
\]
\end{definition}

\subsection{Parsing Algorithms}

We wish to compare the essential characteristics of several parsing
algorithms without getting distracted by implementation details.
\citet{Sikkel1998} introduces a high-level formalism for presenting
and comparing parsing algorithms, called \emph{parsing schemata}, that
present each algorithm as a deductive system. We loosely follow his
approach, but make some changes to better suit our needs.

Each parsing algorithm corresponds to a deductive
system with judgments of the form
\[
  H \vdash \xi 
\]
where $\xi$ is an \emph{item} and $H$ is a set of items.  An item has
the form $[p, i, j]$ where $p$ is either a parse tree or a partial
parse tree and the integers $i$ and $j$ mark the left and right
extents of what has been parsed so far.  The set of \emph{partial parse
  trees} is defined by the following rule.
\begin{quote}
  If $A \to \alpha\beta\gamma \in \mathcal{P}$, then $A \to \alpha \rdot \overline{t}_\beta \rdot
  \gamma$ is a partial parse tree labeled with $A$.
\end{quote}
We reserve the variables $s$ and $t$ for parse trees, not
partial parse trees.  A \emph{complete parse} of an input $w$ of
length $n$ is a derivation of $H_0 \vdash [t_S, 0, n]$, where $H_0$ is
the initial set of items that represent the result of tokenizing the
input $w$.
  \[
  H_0 = \{ [w_i,i,i+1] \mid 0 \leq i < |w| \}
  \]

\begin{example}
  The (top-down) Earley algorithm~\cite{Earley1968,Earley1970} applied
  to a grammar $\G = (\Sigma,\Delta,\mathcal{P},S)$ is defined by the following
  deductive rules.
  \begin{gather*}
    \inference[\textsc{(Hyp)}]{
      \xi \in H
    }{
      H \vdash \xi 
    }
    \quad
    \inference[\textsc{(Fnsh)}]{
      H \vdash [A \to \rdot \overline{t}_\alpha \rdot, i, j]
    }{
      H \vdash [A \to \overline{t}_\alpha, i, j]
    }
    \\[2ex]
    \inference[\textsc{(Init)}]{
      S \to \gamma \in \mathcal{P}
    }{
      H \vdash [S \to \rdot \rdot \gamma, 0, 0]
    }
    \\[2ex]
    \inference[\textsc{(Pred)}]{
      H \vdash [A \to \rdot \overline{t}_\alpha \rdot B \beta,i,j] &
      B \to \gamma \in \mathcal{P}
    }{
      H \vdash [B \to \rdot \rdot \gamma, j, j]
    }
    \\[2ex]
    \inference[\textsc{(Compl)}]{
      H \vdash [A \to \rdot \overline{s}_\alpha \rdot  X \beta,i,j] &
      H \vdash [t_X,j,k] 
    }{
      H \vdash [A \to \rdot \overline{s}_\alpha t_X \rdot \beta, i, k] \} 
    }
  \end{gather*}
\end{example}

\begin{example}
  A bottom-up variation~\cite{Sikkel1998} of Earley parsing is
  obtained by replacing the initialization \textsc{(Init)} and
  prediction \textsc{(Pred)} rules with the following bottom-up rule
  \textsc{(BU)}.
  \begin{gather*}
    \inference[\textsc{(BU)}]{
      H \vdash [t_X, i, j]  &
      A \to X \beta \in \mathcal{P}
    }{
      H \vdash [A \to  \rdot t_X \rdot \beta, i, j]
    }
  \end{gather*}
\end{example}

\section{Type-Oriented Island Parsing}
\label{sec:type-island}

The essential ingredients of our parsing algorithm are type-based
disambiguation and island parsing.  In Section~\ref{sec:evaluation}.
we show that an algorithm based on these two ideas parses with time
complexity that is independent of the number of DSLs in use, so long
as the nonterminals of the DSLs are largely disjoint.  (We also make
this claim more precise.) But first, in this section we introduce our
type-oriented island parsing algorithm (TIP).

Island parsing~\cite{Stock1988} is a bidirectional, bottom-up parsing
algorithm that was developed in the context of speech recognition.  In
that domain, some tokens can be identified with a higher confidence
than others. The idea of island parsing is to begin the parsing
process at the high confidence tokens, the so-called islands, and
expand the parse trees outward from there.

Our main insight is that if our parser can be made aware of variable
declarations, and if a variable's type corresponds to a non-terminal
in the grammar, then each occurrence of a variable is treated as an
island. We introduce the following special form for declaring a
variable $a$ of type $A$ that may be referred to inside the curly
brackets.
\[
  \mathbf{declare}\; a : A\; \{ \ldots \} 
\]
For the purposes of parsing, the rule $A \to a$ is added to the
grammar while parsing inside the curly brackets. To enable temporarily
extending the grammar, we augment the judgments of our deductive
system with an explicit parameter for the grammar. So judgments
have the form
\[
  \G; H \vdash \xi 
\]
This adjustment also enables the import of grammars from different
modules.

We formalize the parsing rule for the $\mathbf{declare}$ form as
follows.
\[
  \inference[\textsc{(Decl)}]{
    \G \cup {(A \to a)}; H \vdash [ t_X, i+5,j]
  }{
    \G; H \vdash [X \to \mathbf{declare}\; a : A\; \{ t_X \}, i,j+1]
  }
\]
Next we split the bottom-up rule \textsc{(BU)} into the two following
rules. The \textsc{(Islnd)} rule triggers the formation of an island
using grammar rules of the form $A \to a$, which arise from variable
declarations and from literals (constants) defined in a DSL.  The
\textsc{(IPred)} rule generates items from grammar rules that have the
parsed nonterminal $B$ on the right-hand side.
\begin{gather*}
  \inference[\textsc{(Islnd)}]{
    \G; H \vdash [a,i,j] &  A \to a \in \mathcal{P} & \G = (\Sigma,\Delta,\mathcal{P},S)
  }{
    \G; H \vdash [A \to a,i,j]
  }
  \\[2ex]
  \inference[\textsc{(IPred)}]{
    \G; H \vdash [t_B, i, j]  \\
    A \to \alpha B \beta \in \mathcal{P} & \G = (\Sigma,\Delta,\mathcal{P},S)
  }{
    \G; H \vdash [A \to  \alpha \rdot t_B \rdot \beta, i, j]
  }
\end{gather*}
Finally, because islands appear in the middle of the input string, we
need both left and right-facing versions of the \textsc{(Compl)} rule.
\begin{gather*}
  \inference[\textsc{(RCompl)}]{
    \G; H \vdash [A \to \alpha \rdot \overline{s}_\beta  \rdot X \gamma,i,j] &
    \G; H \vdash  [t_X,j,k] 
  }{
    \G; H \vdash [A \to \alpha \rdot \overline{s}_\beta t_X \rdot \gamma,i,k] \} 
  }
  \\[2ex]
  \inference[\textsc{(LCompl)}]{
    \G;H \vdash [t_X,i,j]  &
    \G;H \vdash [A \to \alpha X \rdot  \overline{s}_\beta \rdot  \gamma,j,k] 
  }{
    \G;H \vdash [A \to \alpha \rdot t_X \overline{s}_\beta \rdot \gamma,i,k] \} 
  }
\end{gather*}

\begin{definition}
  The \emph{type-oriented island parsing} algorithm is defined 
  as the deductive system comprised of the rules \textsc{(Hyp)},
  \textsc{(Fnsh)}, \textsc{(Decl)}, \textsc{(Islnd)}, \textsc{(IPred)}
  \textsc{(RCompl)}, and \textsc{(LCompl)}.
\end{definition}

The type-oriented island parsing algorithm requires a minor
restriction on grammars. If the right-hand side of a rule does not
contain any nonterminals, then it may only contain a single terminal.
This restriction means that our system supports single-token literals
but not multi-token literals. For example, the grammar rule $A \to
\texttt{"foo"}\; \texttt{"bar"}$ is not allowed, but $A \to
\texttt{"foobar"}$ and $A \to \texttt{"foo"}\;B\; \texttt{"bar"}\;
\texttt{"baz"}$ are allowed.



\section{Experimental Evaluation}
\label{sec:evaluation}

In this section we evaluate the performance of type-oriented island
parsing with experiments in two separate dimensions. First we measure
the performance of the algorithm for programs that are held constant
but the size of the grammars increase, and second we measure the
performance for programs that increase in size while the grammars
are held constant.

\subsection{Grammar Scaling}

\begin{figure*}
  \begin{center}
    \subfigure[normal][Untyped Grammar]{\label{untyped}
    \includegraphics[scale=0.28]{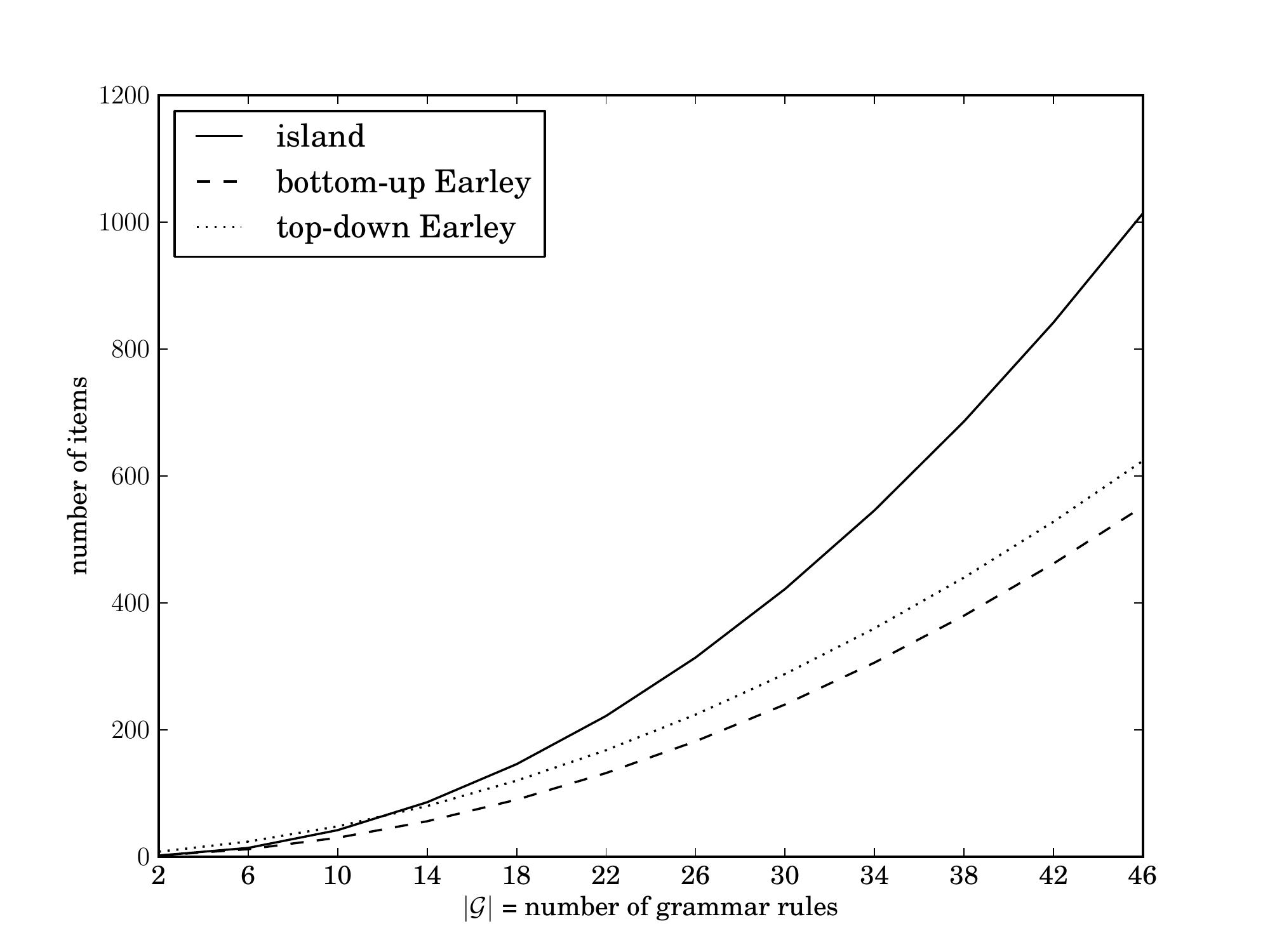}}
    \subfigure[normal][Semi-typed Grammar]{\label{semityped}
    \includegraphics[scale=0.28]{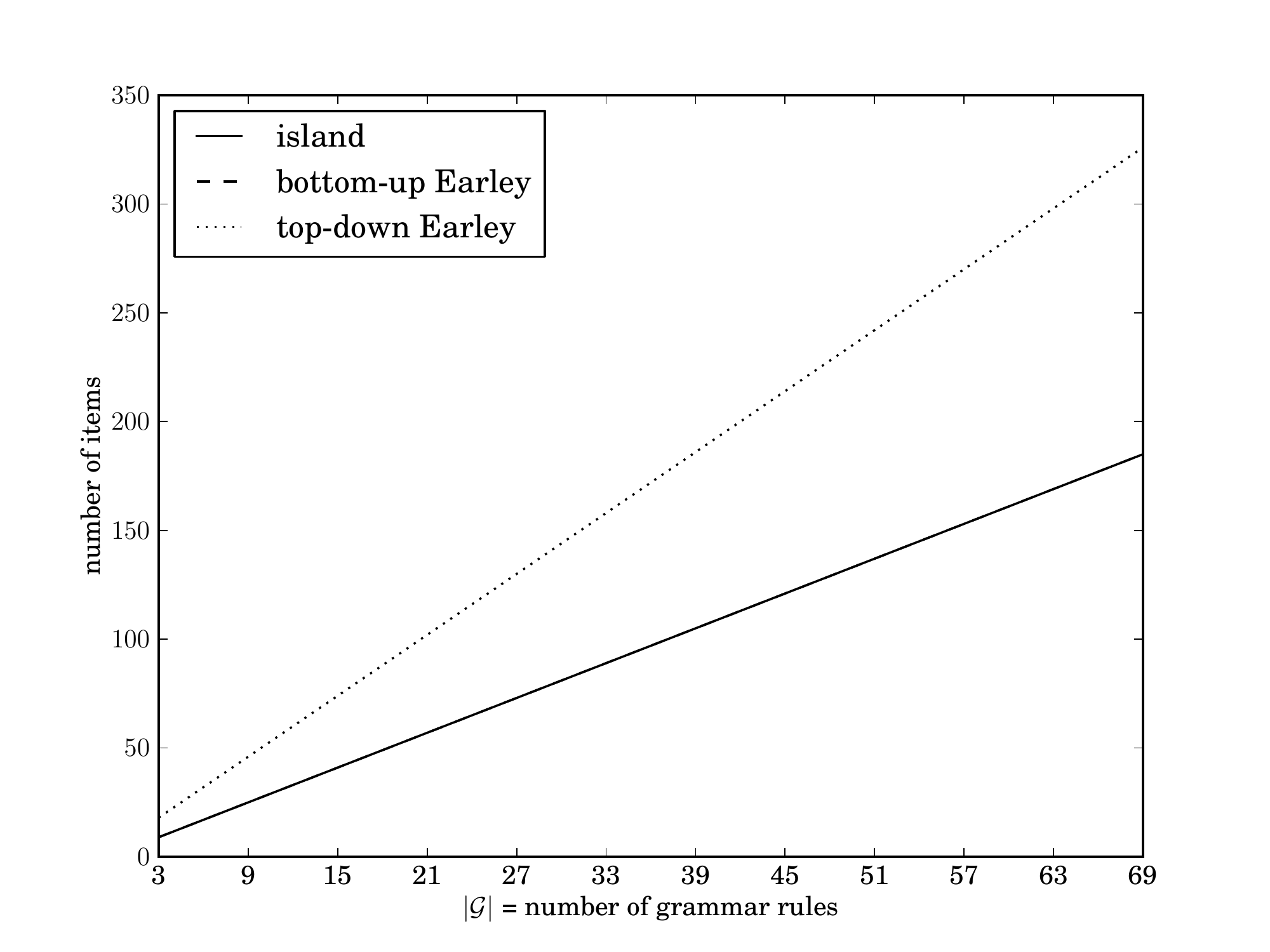}}
    \subfigure[normal][Type-oriented Grammar]{\label{typed}
    \includegraphics[scale=0.28]{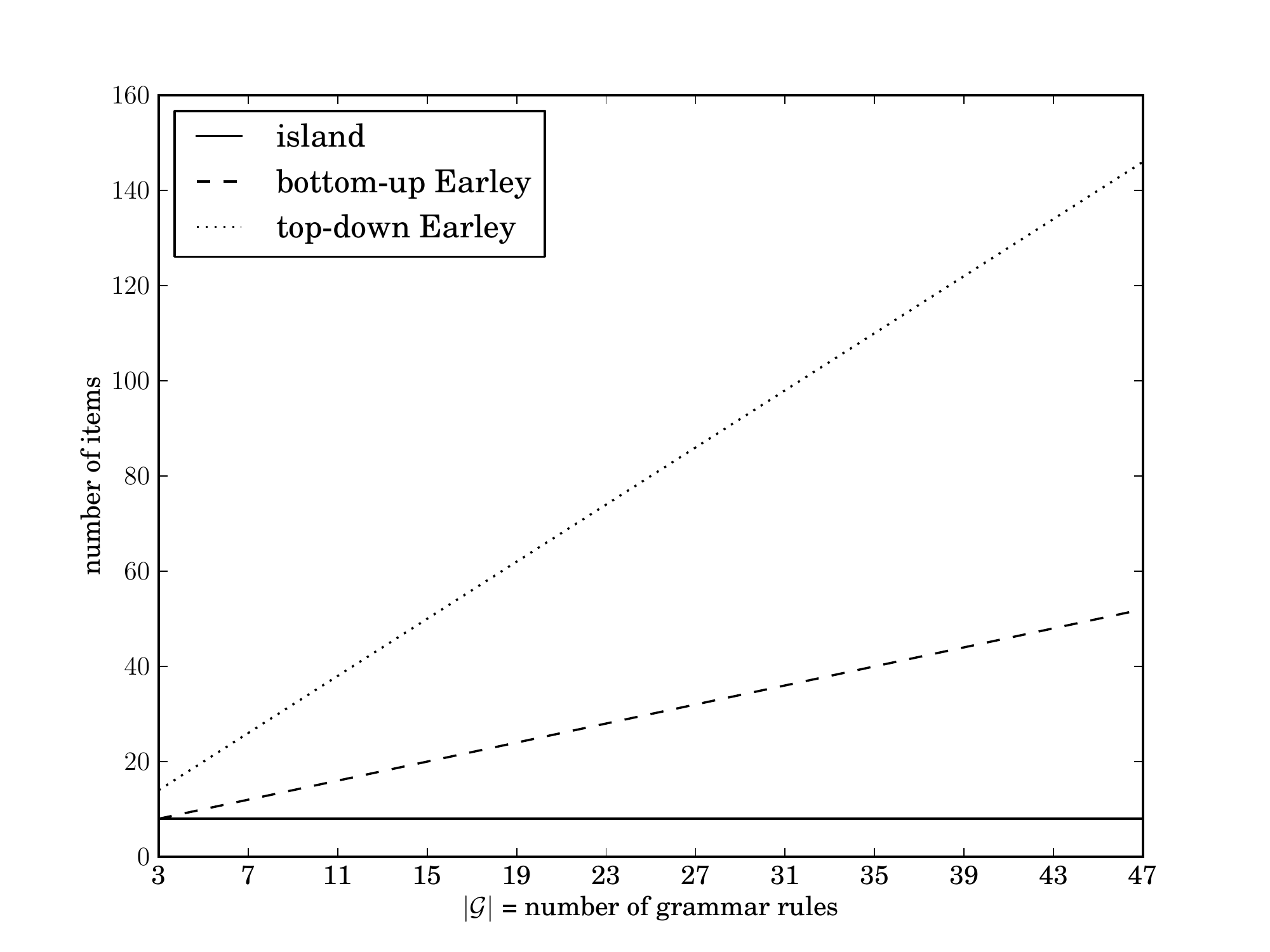}}
  \end{center}
  \caption{Comparison of top-down, bottom-up, and island parsing
    with three styles of grammars.}
  \label{fig:exp-results2}
\end{figure*}

Chart parsing algorithms~\cite{Kay1986} have a general worst-case running
time of $O(|\G|n^3)$ for a grammar \G{} and string of length $n$.  
In our setting, \G{} is the union of the grammars for all the $k$ DSLs
that are in use within a given scope, that is $\G = \bigcup_{i=1}^k \G_i$, where
$\G_i$ is the grammar for DSL $i$.
We claim that the total size of the grammar \G{} is not a factor for
type-oriented island parsing, and instead the time complexity is $O(m
n^3)$ where $m = \mathrm{max} \{ |\G_i| \mid 1 \leq i \leq k \}$.  This claim deserves considerable explanation to
be made precise.

Technically, we assume that \G{} is \emph{sparse} and that the terminals
of \G{} are \emph{well-typed}, which we define as follows.
\begin{definition}
  Form a Boolean matrix with a row for each nonterminal
  and a column for each production rule in a grammar \G{}. A matrix
  element $(i,j)$ is \textsf{true} if the nonterminal $i$ appears on the
  right-hand side of the rule $j$, and it is \textsf{false} otherwise. We
  say that \G{} is \emph{sparse} if its corresponding matrix is sparse,
  that is, if the number of nonzero elements is much smaller than the
  number of elements.
\end{definition}
\begin{definition}
  We say that a terminal $a$ of a grammar \G{} is \emph{well-typed} if
  for each $B$ such that $B \to a \in \mathcal{P}$, $B$ represents a
  type in the language of \G{}.
\end{definition}
We expect that the terminals of type-oriented grammars will be
well-typed, and hypothesize that, in the common case, the union of many
type-oriented grammars (or DSLs) is sparse.

To verify that both the type-oriented style of grammar and the island
parsing algorithm are necessary for this result, we show that removing
either of these ingredients results in parse times that are dependent on
the size of the entire grammar. Specifically, we consider the performance
of the top-down and bottom-up Earley algorithms, in addition to island
parsing, with respect to untyped, semi-typed, and 
type-oriented grammars.

We implemented all three algorithms in a chart parsing
framework~\cite{Kay1986}, which efficiently memoizes duplicate items.
The chart parser continues until it has generated all items that can
be derived from the input string. (It does not stop at the first
complete parse because it needs to continue to check whether the input
string is ambiguous, which means the input would be in error.)
Also, we should note that our system currently employs a
fixed tokenizer, but that we plan to look into scannerless parsing.



To capture the essential, asymptotic behavior of the parsing
algorithms, we measure the number of items generated during the
parsing of the program.


%



\subsubsection{A Small Experiment}

For the first experiment we parse the expression \texttt{--A} with
untyped, semi-typed, and typed grammars.

\paragraph{Untyped} In the untyped scenario, all grammar rules are defined
in terms of the expression nonterminal ({\tt E}), and variables are simply
parsed as identifiers ({\tt Id}).
\begin{alltt}
  {\bf{module}} Untyped\(\sp{k}\) \{
    E ::= Id | "-" E;
  \}
\end{alltt}

The results for parsing \texttt{--A} after importing $k$ copies of
\texttt{Untyped}, for increasing $k$, are shown in Figure~\ref{untyped}.
The y-axis is the number of items generated by each parsing algorithm,
and the x-axis is the total number of grammar rules at each $k$.
In the untyped scenario, the size of the grammar affects the performance
of each algorithm, with each generating $O(k^2)$ items.

We note that the two Earley algorithms generate about half as many items as
the island parser because they are unidirectional (left-to-right)
instead of bidirectional.

\paragraph{Semi-typed} In the semi-typed scenario, the grammars are
nearly type-oriented: the $\texttt{Semityped}^0$ module defines
the nonterminal {\tt V} (for vector) and each of $\texttt{Semityped}^i$
for $i \in \{1,\ldots,k\}$ defines the nonterminal $\texttt{M}i$
(for matrix); however, variables are again parsed as identifiers.
We call this scenario \emph{semi-typed}, because it doesn't use variable
declarations to provide type-based disambiguation.
\begin{alltt}
  {\bf{module}} Semityped\(\sp{0}\) \{
    E ::= V;
    V ::= Id | "-" V;
  \}
  {\bf{module}} Semityped\(\sp{i}\) \{
    E ::= M\(i\);
    M\(i\) ::= Id | "-" M\(i\);
  \}
\end{alltt}

The results for parsing \texttt{--A} after importing
$\texttt{Semityped}^0$ followed by $\texttt{Semityped}^i$ for $i \in
\{1,\ldots,k\}$ are shown in Figure~\ref{semityped}.  The lines for
bottom-up Earley and island parsing coincide.  Each algorithm
generates $O(k)$ items, but we see that type-oriented grammars are
not, by themselves, enough to achieve constant scaling with respect to
grammar size.

We note that the top-down Earley algorithm generates almost twice as many
items as the bottom-up algorithms: the alternatives for the start symbol
{\tt E} grow with $n$, which affects the top-down strategy more than
bottom-up.

\paragraph{Typed} The typed scenario is identical to semi-typed except
that it no longer includes the {\tt Id} nonterminal. Instead, programs
using the {\tt Typed} module must declare their own typed variables.
\begin{alltt}
  {\bf{module}} Typed\(\sp{0}\) \{
    E ::= V;
    V ::= "-" V;
  \}
  {\bf{module}} Typed\(\sp{i}\) \{
    E ::= M\(i\);
    M\(i\) ::= "-" M\(i\);
  \}
\end{alltt}

The results for parsing \texttt{--A} after importing $\texttt{Typed}^0$
followed by $\texttt{Typed}^i$ for $i \in \{1,\ldots,k\}$ and declaring
{\tt A:V} are shown in Figure~\ref{typed}.
The sparsity for this example is $O(1/k)$, and now the terminal
(as {\tt V}) is well-typed.
The island parsing algorithm generates a constant number of items as the
size of the grammar increases, while the Earley algorithms remain linear.
Thus, 
the \emph{combination} of type-based disambiguation, type-oriented
grammars, and island parsing provides a scalable approach to parsing
programs that use many DSLs.




\subsubsection{A Larger Experiment}



\begin{figure}
  \centering
  \includegraphics[scale=0.4]{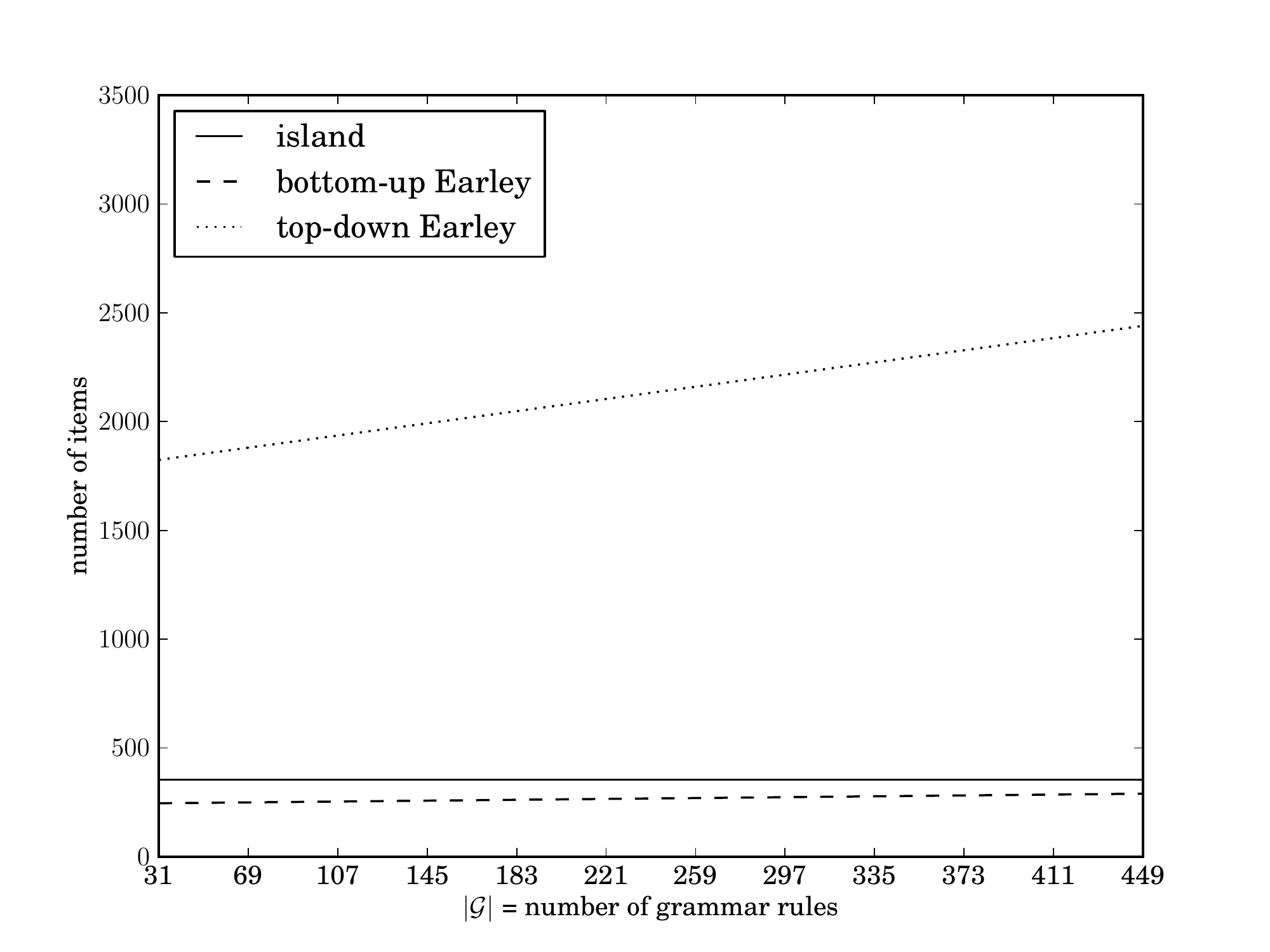}
  \caption{Comparison of parsing algorithms
    for a type-oriented matrix algebra DSL and increasing grammar size.}
  \label{fig:exp-results1}
\end{figure}

For a second experiment we measure the performance of each algorithm
for a sequence of matrix algebra operations with expanded versions of
the grammars in Figure~\ref{fig:composing-disamb}:
\begin{alltt}
  {\bf{import}} MatrixAlgebra, RegularExpressions\(\sp{k}\), Sets\(\sp{k}\);
  B = A + u1 * v1' + u2 * v2';
  x = b * (B' * y) + z;
  w = a * (B * x);
\end{alltt}
In this example, we import grammars for {\tt RegularExpressions} and
{\tt Sets}, $k$ times each. For the untyped and semi-typed scenarios,
the result is too ambiguous and we terminated their execution after
waiting for several minutes. For the typed scenario, we declare the
variables {\tt A} and {\tt B} as type {\tt Matrix}; {\tt u1}, {\tt
  u2}, {\tt v1}, {\tt v2}, and {\tt w}-{\tt z} as type {\tt
  ColVector}; {\tt a} and {\tt b} as type {\tt Scalar}; the sparsity
of the typed example is $O(1/k)$.

Figure~\ref{fig:exp-results1} shows a graph for parsing the above program
with each algorithm. As before, the y-axis is the number of items generated
during parsing, and the x-axis is the number of DSLs that are imported.
The top-down Earley algorithm scales linearly with respect to the
number of DSLs imported and generates many more items than the
bottom-up algorithms. The island parsing algorithm generates a
constant number of items as the number of DSLs increases; the
bottom-up Earley algorithm generates a similar number of items, but it
scales slightly linearly.

\subsubsection{Discussion}

The reason type-oriented island parsing scales is that it is more
conservative with respect to prediction than either top-down or bottom
up, and so grammar rules from other DSLs that are irrelevant
to the program fragment being parsed are never used to generate items.

Consider the \textsc{(Pred)} rule of top-down Earley parsing.  Any
rule that produces the non-terminal $B$, regardless of which DSL it
resides in, will be entered into the chart. Note that such items have
a zero-length extent which indicates that the algorithm does not yet
have a reason to believe that this item will be able to complete.

Looking at the \textsc{(BU)} rule of bottom-up Earley parsing, we see
that all it takes for a rule (from any DSL) to be used is that it
starts with a terminal that occurs in the program. However, it is quite
likely that different DSLs will have rules with some terminals in common.
Thus, the bottom-up algorithm also introduces items from irrelevant
DSLs.

Next, consider the \textsc{(Islnd)} rule of our island parser. There
is no prediction in this rule. However, it is possible for different
DSLs to define literals with the same syntax (same
tokens). (Many languages forbid the overloading of constants, but it
is allowed, for example, in Haskell.) The performance of the island
parser would degrade in such a scenario, although the programmer could
regain performance by redefining the syntax of the imported constants,
in the same way that name conflicts can be avoided by the
rename-on-import constructs provided by module systems.

Finally, consider the \textsc{(IPred)} rule of our island parser.  The
difference between this rule and \textsc{(BU)} is that it only applies
to nonterminals, not terminals. As we previously stated, we assume
that the nonterminals in the different DSLs are, for the most part,
disjoint. Thus, the \textsc{(IPred)} rule typically generates items
based on rules in the relevant DSL's grammar and not from other DSLs.


\subsection{Program Scaling}

\begin{figure*}
  \begin{center}
    \subfigure[normal][Untyped Grammar]{\label{untyped-ps}
    \includegraphics[scale=0.28]{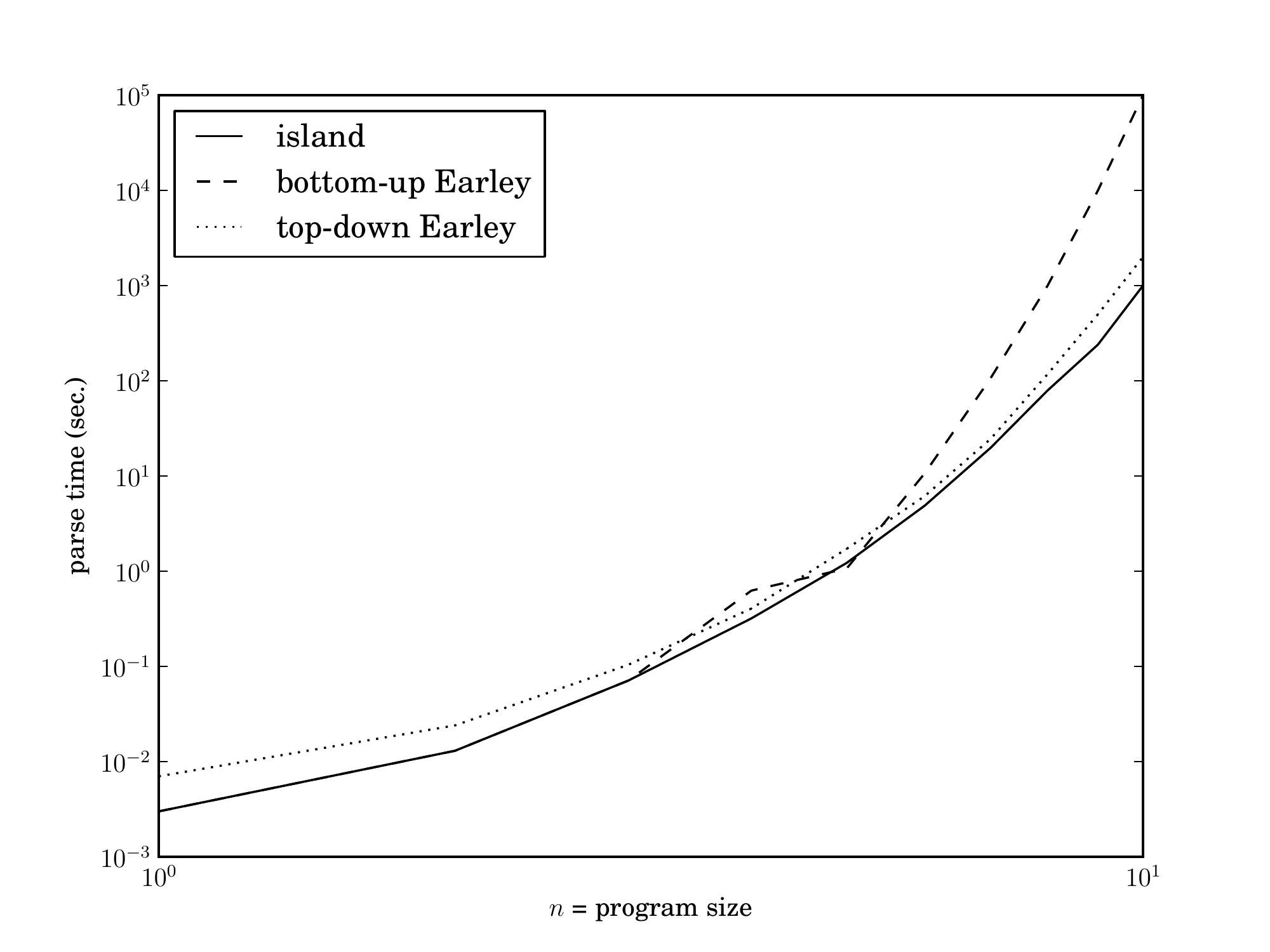}}
    \subfigure[normal][Semi-typed Grammar]{\label{semityped-ps}
    \includegraphics[scale=0.28]{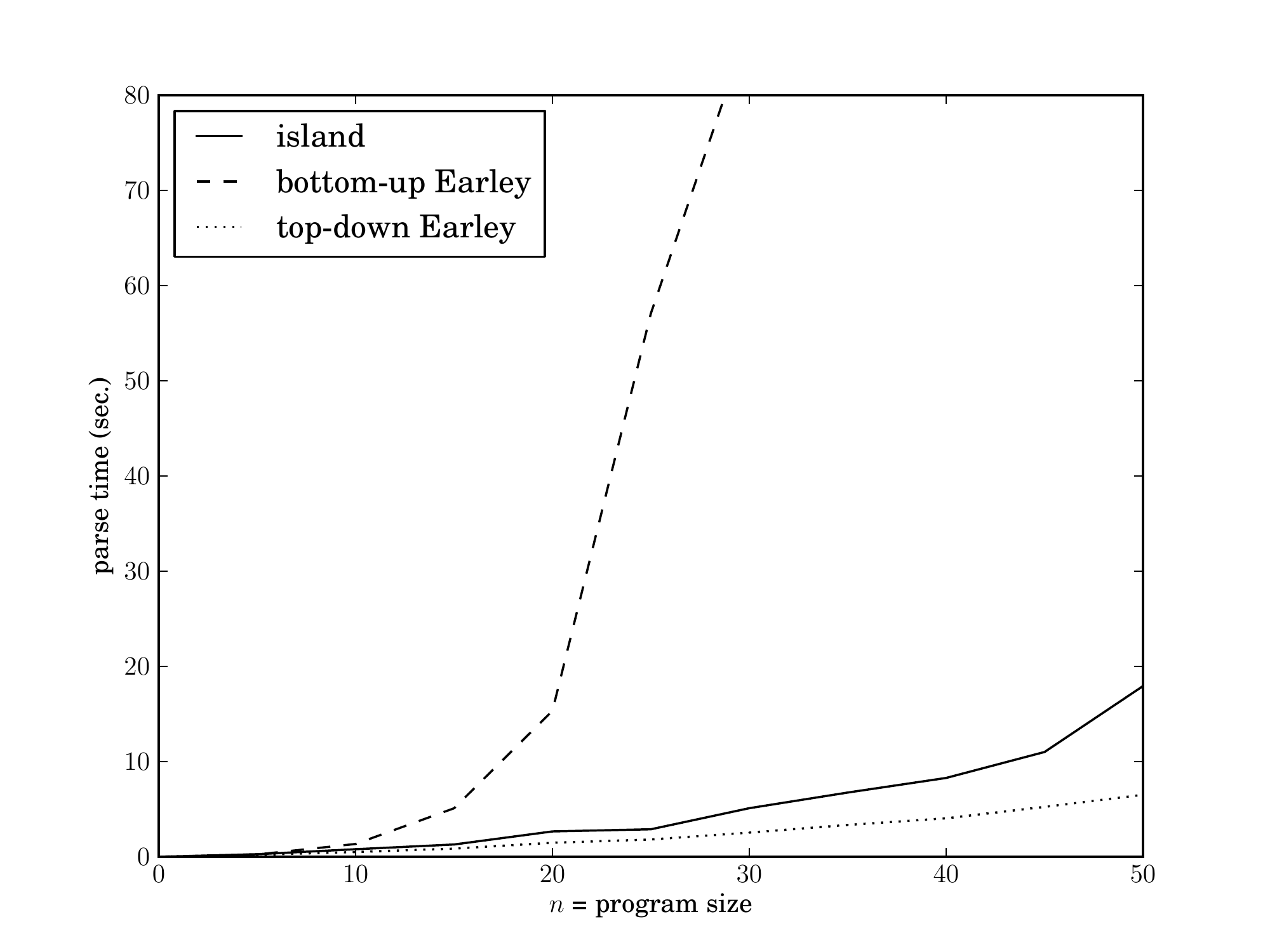}}
    \subfigure[normal][Type-oriented Grammar]{\label{typed-ps}
    \includegraphics[scale=0.28]{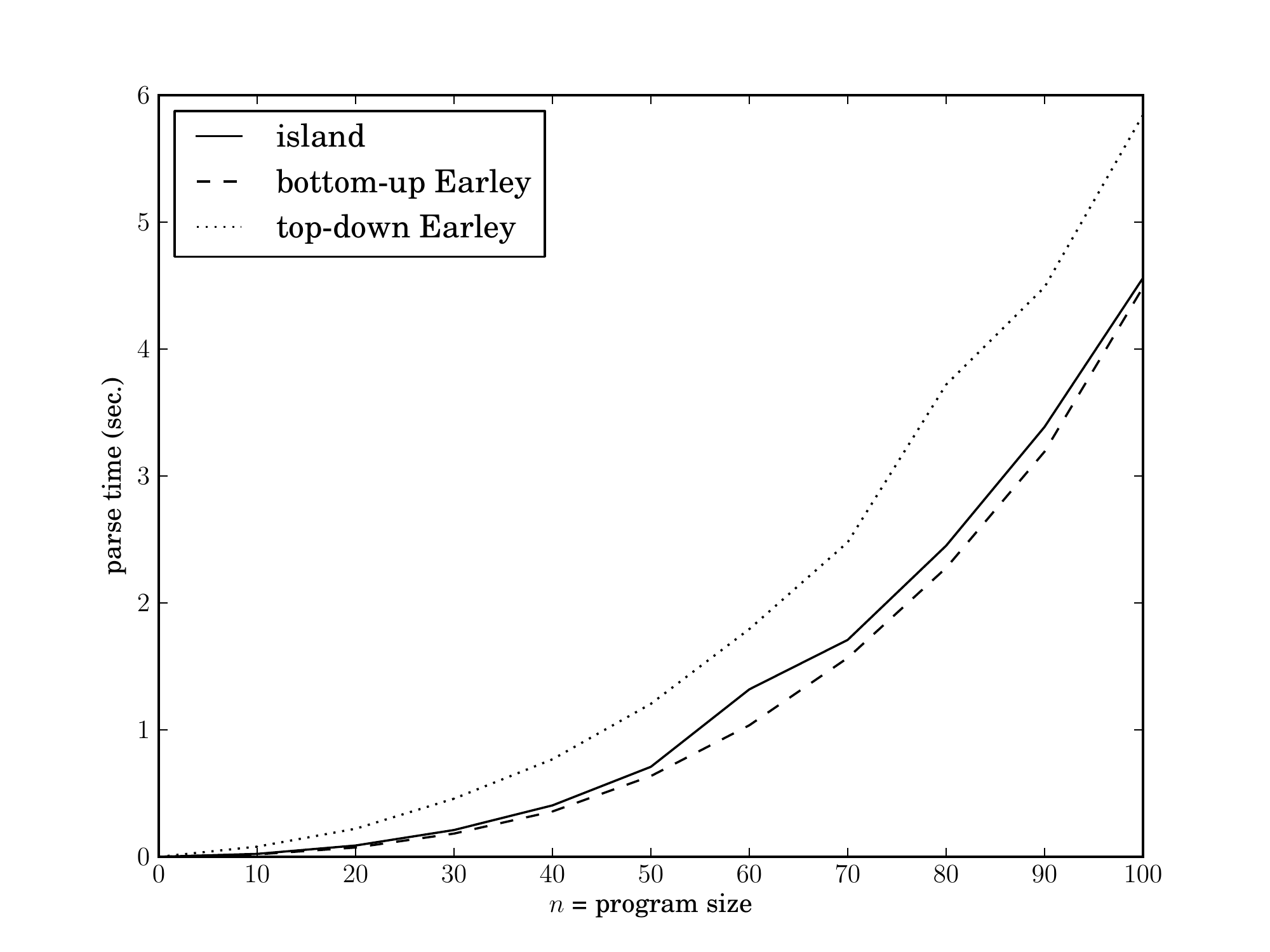}}
  \end{center}
  \caption{Comparison of parsing algorithms for increasing program
    size.  Figure (a) uses a logarithmic scale, with the program size
    ranging from 1 to 10. Figures (b) and (c) use a linear scale, with
    the program size ranging from 1 to 50 in (b) and ranging from 1 to
    100 in (c).}
  \label{fig:ps-results2}
\end{figure*}


In this section we measure the performance of each algorithm as the size of
the program increases and the grammar is held constant. The program of
size $n$ is the addition of $n$ matrices using the matrix algebra grammar
from the previous section.

As before, we consider untyped, semi-typed, and typed scenarios. For these
experiments we report parse times; we ran all of the experiments on a
MacBook with a 2.16 GHz Intel Core 2 Duo processor and 2 GB of RAM.

\paragraph{Untyped} The untyped scenario is exponentially ambiguous: 
\begin{alltt}
  {\bf import} MatrixAlgebra, RegularExpressions, Sets;
  A + A + \(\cdots\) + A
\end{alltt}
While the above program with $n$ terms produces $O(2^n)$ parse trees,
the Earley and island parsing algorithms can produce a packed parse
forest in polynomial space and time~\cite{Allen:1995}.

Figure~\ref{untyped-ps} shows the results for each algorithm on a
logarithmic scale. The y-axis is the parse time (including production of
parse trees), and the x-axis is the program size. Because our implementation
does not use packed forests, all three algorithms
are exponential for the untyped scenario.

\paragraph{Semi-typed} The program for the semi-typed scenario is
identical to the untyped scenario and is also ambiguous; however, the
number of parse trees doesn't grow with increasing program size.
Figure~\ref{semityped-ps} shows the results for each algorithm, now on
a linear scale. The axes are the same as before. Here the top-down
Earley and island algorithms are $O(n^2)$. Although the number of
correct parse trees remains constant, the bottom-up Earley algorithm
explores an exponential number of possible trees as $n$ increases
before returning, and uses exponential time.

\paragraph{Typed} The program is no longer ambiguous in the typed scenario:
\begin{alltt}
  {\bf{import}} MatrixAlgebra, RegularExpressions, Sets;
  {\bf{declare}} A:Matrix \{
    A + A + \(\cdots\) + A
  \}
\end{alltt}
Figure~\ref{typed-ps} shows the results for each algorithm on
a linear scale and with axes as before. All three algorithms are $O(n^2)$
for the typed scenario. These results suggest that type-oriented Earley
and island parsing are $O(n^2)$ for unambiguous grammars.

We should note that the top-down Earley algorithm parses the above program in
$O(n)$ time when the grammars are rewritten to be LR(0); however, the
bottom-up Earley and island algorithms remain $O(n^2)$.







\section{A System for Extensible Syntax}
\label{sec:bells-whistles}


In this section we describe the parsing system that we have built as a
front end to the Racket programming language. In particular, we
describe how we implement four features that are needed in a practical
extensible parsing system: associativity and precedence, parameterized
grammar rules, grammar rules with variable binders and scope, and
rule-action pairs~\cite{Sandberg1982} which combine of the notions of
semantic actions, function definitions, and macros.

\subsection{Associativity and Precedence}

We view associativity and precedence annotations (as in
Figure~\ref{composing-amb}, e.g., {\tt [left,1]}) as a must for our
parsing system because we do not expect all of our users to be
computer scientists, that is, we do not expect them to know how to
manually factor a grammar to take associativity and precedence into
account. Further, even for users who are computer scientists, they
probably have something better to do with their time than to factor
grammars.

Our treatment of associativity and precedence is largely based on
that of \citet{Visser1997}, although we treat this as a semantic
issue instead of an optimization issue.
From the user perspective, we extend rules to have the form $A \to \alpha
[\ell,p]$ where $\ell$ indicates the associativity, where $\ell \in
\{\textsf{left},\textsf{right},\textsf{non}, \bot\}$, and $p$ indicates
the precedence, where $p \in \mathbb{N}_{\bot}$. We use an ordering $<$ on
precedences that is the natural lifting of $<$ on $\mathbb{N}$.
(Instead of $(\mathbb{N}_{\bot},<)$ we could use any partially ordered
set, but prefer to be concrete here.)

To specify the semantics of precedence and associativity, we use the
notion of a \emph{filter} to remove parse trees from consideration if
they contain precedence or associativity
conflicts~\cite{Visser1997}. But first, we annotate our parse trees
with the precedence and associativity, so an internal node has the
form $\htree{A}{\overline{t}}{\ell,p}$.

\begin{definition}
  We say that a parse tree $t$ has a \emph{root priority conflict},
  written $\mathit{conflict}(t)$, if one of the following holds.
  \begin{enumerate}
  \item It violates the right, left or non-associativity rules, that is,
    $t$ has the form:
    \begin{itemize}
    \item $\htree{A}{(\htree{A}{\overline{t}_{A\alpha}}{\ell,p}) \overline{s}_\alpha}{\ell,p}$
      where $\ell = \mathsf{right}$  or $\ell = \mathsf{non}$.
    \item $\htree{A}{\overline{s}_\alpha (\htree{A}{\overline{t}_{\alpha A}}{\mathsf{\ell},p}) }{\mathsf{\ell},p}$ where
      $\ell = \mathsf{left}$ or $\ell = \mathsf{non}$.
    \end{itemize}
  \item It violates the precedence rule, that is, $t$ has the form:
    \[
    t = \htree{A}{\overline{s} (\htree{B}{\overline{t}}{\ell',p'}) \overline{s'} }{\ell,p}
    \text{ where } p' < p.
    \]
  \end{enumerate}
\end{definition}

\begin{definition}
  A \emph{tree context} $\mathcal{C}$ is defined by the following
  grammar.
\[
\begin{array}{rcl}
  \C{} & ::= & \Box \mid \htree{A}{t_1\ldots\C{}\ldots t_n}{l,p}
\end{array}
\]
The operation of plugging a tree $t$ into a tree context \C{}, written
$\C[t]$, is defined as follows.
\begin{align*}
  \Box[t] &= t \\
  (\htree{A}{t_1\ldots\C{}\ldots t_n}{\ell,p})[t] &= \htree{A}{t_1\ldots\C{}[t]\ldots t_n}{\ell,p}
\end{align*}
\end{definition}

\begin{definition}
  The \emph{filter} for a CFG \G{} is a function on sets of trees,
  $\mathcal{F}\colon \wp(\ptrees) \to \wp(\ptrees)$, that removes the
  trees containing conflicts. That is,
  \[
  \mathcal{F}(\Phi) = \{ t \in \Phi \mid \not \exists t'\mathcal{C},\, t = \mathcal{C}[t'] 
                              \text{ and } \mathit{conflict}(t') \}
  \]
\end{definition}

\begin{definition}
  The set of parse trees for a grammar \G{} (with precedence and
  associativity) and input $w$, written $\mathcal{T}(\G,w)$, is
  defined as follows.
  \[
  \mathcal{T}(\G,w) = \{ t_S \mid t_S \in \filter{\mathcal{T}(\G)} 
     \text{ and } \mathit{yield}(t_S) = w \} 
  \]
\end{definition}

The change to the island parsing algorithm to handle precedence and
associativity is straightforward. We simply make sure that a partial
parse tree does not have a root priority conflict before converting it
into a (complete) parse tree. We replace the \textsc{(Fnsh)} rule with
the following rule.
\[
\inference[\textsc{(FnshP)}]{
  \G; H \vdash [A \to \rdot \overline{t}_\alpha \rdot, i, j] &
  \lnot \mathit{conflict}(A \to \overline{t}_\alpha)
}{
  \G; H \vdash [A \to \overline{t}_\alpha, i, j]
}
\]

\subsection{Parameterized Rules}

With the move to type-oriented grammars, the need for parameterized
rules immediately arises. For example, consider how one might
translate the following grammar rule for conditional expressions into
a type-oriented rule.
\begin{alltt}
  E ::= "if" E "then" E "else" E
\end{alltt}
We would like to be more specific than {\tt E} for the two branches
and for the left-hand side. So we extend our grammar
rules to enable the parameterization of nonterminals. We can
express a conditional expression as follows, where {\tt T} stands for
any type/nonterminal.
\begin{alltt}
  forall T.
    T ::= "if" Bool "then" T "else" T
\end{alltt}

To simplify the presentation, we describe parameterized rules as an
extension to the base island parser (without precedence).  However,
our parsing system combines both extensions.  Here we extend
grammar rules to include parameters: $\forall \overline{x}.\, A \to \alpha$.
($\overline{x}$ may not contain duplicates.) We use $x,y,z$ to range over
variables and we now use the variables $A,B,C,D$ to range over
nonterminals and variables. 

To handle parameters we need the notion of a substitution $\sigma$, that
is, a partial function from variables to nonterminals. The initial
substitution $\sigma_0$ is everywhere undefined.  We extend the action of
a substitution to all symbols, sequences, and rules in the following
natural way.
\begin{align*}
  \sigma(a) &= a \\
  \sigma(X_1 \ldots X_n) &= \sigma(X_1) \ldots \sigma(X_n) \\
  \sigma(A \to \alpha) &= \sigma(A) \to \sigma(\alpha)
\end{align*}
The
notation $\sigma[X \mapsto Y]$ creates a new, extended substitution,
defined as follows.
\[
\sigma[X \mapsto Y](Z) =
\begin{cases}
  Y & \text{if } X = Y, \\
  \sigma(Z) & \text{otherwise}.
\end{cases}
\]
We write $\sigma[\overline{X} \mapsto \overline{Y}]$ to abbreviate
\[\sigma[X_1 \mapsto Y_1]\cdots[X_{|X|} \mapsto Y_{|Y|}].\]
We write $[\overline{X} \mapsto \overline{Y}]$ to abbreviate
$\sigma_0[\overline{X} \mapsto \overline{Y}]$.

Next we update the definition of a parse tree to include parameterized
rules. The formation rule for leaves remains unchanged, but the rule
for internal nodes becomes as follows.
\begin{quote}
  If $A \in \Delta$, $\forall \overline{x}.\, A \to \alpha \in
  \mathcal{P}$, and $\sigma = [\overline{x}\mapsto \overline{B}]$, then
  $\sigma(A) \to \overline{t}_{\sigma(\alpha)}$ is a parse tree
  labeled with $\sigma(A)$.
\end{quote}

The definition of the language of a CFG with parameterized rules
requires some care because parameterized rules introduce ambiguity.
For example, consider the parameterized rule for conditional
expressions given above and the following program.
\begin{alltt}
  if true then 0 else 1
\end{alltt}
Instantiating parameter \texttt{T} with either \texttt{Int} or
\texttt{E} leads to a complete parse. Of course, instantiating with
\texttt{Int} is better in that it is more specific. We formalize this
notion as follows.

\begin{definition}
  We inductively define whether $A$ \emph{is at least as specific
    as} $B$, written $A \geq B$, as follows.
  \begin{enumerate}
  \item If $B \to A \in \mathcal{P}$, then $A \geq B$.
  \item (reflexive) $A \geq A$.
  \item (transitive) If $A \geq B$ and $B \geq C$, then $A \geq C$.
  \end{enumerate}
  We extend this ordering to terminals by defining $a \geq b$ iff $a = b$,
  and to sequences by defining
  \[
    \alpha \geq \beta \text{ iff } |\alpha| = |\beta| \text{ and } 
      \alpha_i \geq \beta_i  \text{ for } i \in \{ 1,\ldots,|\alpha| \}
  \]
  A parse tree node $A \to \overline{s}_\alpha$ is at least as specific as
  another parse tree node $B \to \overline{t}_\beta$ if and only if $A \geq 
  B$ and $\overline{s}_\alpha \geq \overline{t}_\beta$.

  We define the least upper bound, $A \lor B$, with respect to the $\geq$
  relation in the usual way. Note that a least upper bound does not
  always exist.
\end{definition}

\begin{definition}
  The \emph{language} of a CFG \G{} with parameterized rules, written
  $L(\mathcal{G})$, consists of all the strings for which there is a
  most specific parse tree. More formally,
  \[
  L(\G) = \{ w \mid \exists t \in \mathcal{T}(\G,w).\, 
    \forall t' \in \mathcal{T}(\G,w).\, t' \neq t \to t \geq t' \} 
  \]
\end{definition}

Next we turn to augmenting our island parsing algorithm to deal with
parameterized rules. We wish to implicitly instantiate parameterized
grammar rules, that is, automatically determine which nonterminals to
substitute in for the parameters. Towards this end, we define a
partial function named $\mathit{match}$ that compares two symbols with
respect to a substitution $\sigma$ and list of variables $\overline{y}$
and produces a new substitution $\sigma'$ (if the match is successful).
\begin{align*}
  \mathit{match}(X, X, \sigma, \overline{y}) & =  \sigma \\
  \mathit{match}(x, Y, \sigma, \overline{y}) & =
  \begin{cases}
    \sigma[x \mapsto X \lor Y] & \text{if } x \in \overline{y} \text{ and}\\
                  &  \sigma(x) = X \\
    \sigma[x \mapsto Y] & \text{if } x \in \overline{y} \text{ and} \\
              &  x \notin \mathrm{dom}(\sigma)
  \end{cases}
\end{align*}

Next, we augment a partial parse tree with a substitution to
incrementally accumulate the matches. So a partial tree has the form
$\forall \overline{x}.\, A \to^\sigma \alpha \rdot \overline{t}_\beta \rdot \gamma$.  We
then update four of the deduction rules as shown below, leaving
\textsc{(Hyp)}, \textsc{(Decl)}, and \textsc{(Islnd)} unchanged.
\begin{gather*}
    \inference[\textsc{(PFnsh)}]{
      \G; H \vdash [\forall \overline{x}.\, A \to^\sigma \rdot \overline{t}_\alpha \rdot, i, j] 
    }{
      \G; H \vdash [\sigma(A) \to \overline{t}_\alpha, i, j]
    }
    \\[2ex]
  \inference[\textsc{(PIPred)}]{
    \G; H \vdash [t_B, i, j]  & \mathit{match}(B', B, \sigma_0, \overline{x}) = \sigma \\
    \forall \overline{x}.\, A \to \alpha B' \beta \in \mathcal{P} & \G = (\Sigma,\Delta,\mathcal{P},S)
  }{
    \G; H \vdash [\forall \overline{x}.\, A \to^\sigma \alpha \rdot t_B \rdot \beta, i, j]
  }
  \\[2ex]
  \inference[\textsc{(PRCompl)}]{
    \G; H \vdash [\forall \overline{x}.\, A \to^{\sigma_1} \alpha \rdot \overline{s}_\beta  \rdot X' \gamma,i,j] \\
    \G; H \vdash  [t_X,j,k] \\
    \mathit{match}(X', X, \sigma_1, \overline{x}) = \sigma_2
  }{
    \G; H \vdash [\forall \overline{x}.\, A \to^{\sigma_2} \alpha \rdot \overline{s}_\beta t_X \rdot \gamma,i,k] \} 
  }
  \\[2ex]
  \inference[\textsc{(PLCompl)}]{
    \G;H \vdash [t_X,i,j]  \\
    \G;H \vdash [\forall \overline{x}.\, A \to^{\sigma_1} \alpha X' \rdot  \overline{s}_\beta \rdot  \gamma,j,k] \\
    \mathit{match}(X', X, \sigma_1, \overline{x}) = \sigma_2
  }{
    \G;H \vdash [\forall \overline{x}.\, A \to^{\sigma_2} \alpha \rdot t_X \overline{s}_\beta \rdot \gamma,i,k] \} 
  }
\end{gather*}

The above rules ensure that we instantiate type parameters in a way
that generates the most specific parses for parameterized rules, but there
is still the possibility of ambiguities in non-parametric rules.  For
example, consider the following grammar.
\begin{alltt}
  Float ::= Int
  Float ::= Float "+" Float
  Int ::= Int "+" Int
\end{alltt}
The program
\begin{alltt}
  1 + 2
\end{alltt}
can be parsed at least three different ways, with no coercions from
\texttt{Int} to \texttt{Float}, with two coercions, or with just one
coercion. To make sure that our algorithm picks the most specific
parse, with no coercions, we make sure to explore derivations in the
order of most specific first.

\subsection{Grammar Rules with Variable Binders}
\label{sec:binders}

Variable binding and scoping is an important aspect of programming
languages and domain-specific languages are no different in this
regard. Consider what would be needed to define the grammar rule to
parse a \texttt{let} expression such as the following, in which
\texttt{n} is in scope between the curly brackets.
\begin{alltt}
  let n = 7 \{ n * n \}
\end{alltt}
To facilitate the definition of binding forms, we add two extensions
to our extensible parsing system: labeled symbols~\cite{Jim:2010ve} and
a scoping construct~\cite{Cardelli1994}. First, to see an example,
consider the below grammar rule.
\begin{alltt}
  forall T1 T2.
    T2 ::= "let" x:Id "=" T1 "\{" x:T1; T2 "\}"
\end{alltt}
The identifier \texttt{Id} is now labeled with \texttt{x}, which
provides a way to refer to the string that was parsed as \texttt{Id}.
The curly brackets are our scoping construct, that is, they are
treated specially. The \texttt{x:T1} inside the curly brackets
declares that \texttt{x} is in scope during the parsing of
\texttt{T2}.  Effectively, the grammar is extended with the rule
$\texttt{T1}\to \texttt{x}$ (but with \texttt{T1} replaced by the
nonterminal that it is instantiated to, and with \texttt{x} replaced
by its associated string).

The addition of variable binders and scoping complicates the parsing
algorithm because we can no longer proceed purely in a bottom-up
fashion. In this example, we cannot parse inside the curly brackets
until we have parsed the header of the \texttt{let} expression, that
is, the variable name and the right-hand side \texttt{T1}.  Our
parsing system handles this by parsing in phases, where initially, all
regions of the input enclosed in curly braces are ignored. Once enough
of the text surrounding a curly-brace enclosed region has been parsed,
then that region is ``opened'' and the next phase of parsing begins.

\subsection{Rule-Action Pairs and Nonterminal-Type Mappings}

\citet{Sandberg1982} introduces the notion of a \emph{rule-action
  pair}, which pairs a grammar rule with a semantic action that
provides code to give semantics to the syntax. The following is one of
his examples but written using our parsing system on top of Typed
Racket.
\begin{alltt}
  Integer ::= "|" i:Integer "|" => (abs i);
\end{alltt}
The above example defines syntax for the absolute value operation on
integers and it defines how to compute the absolute value with code in
Typed Racket. After a declaration such as the one above, the
programmer can use the notation \texttt{|x|} in the rest of the
current scope, including subsequent actions within rule-action pairs.

In Sandberg's paper, it seems that rule-action pairs behave like
macros. In our system, we provide rule-action pairs that behave like
functions as well (with call-by-value semantics). The \texttt{=>}
operator introduces a function (as in the above example) and the
\texttt{=} operator introduces a macro. For example, one would want to
use a macro to define the syntax of an \texttt{if} expression
(Figure~\ref{fig:let-example}) to avoid always evaluating both
branches.  We refer to a rule-action pair that defines a function as a
\emph{rule-function} and a rule-action pair that defines a macro as a
\emph{rule-macro}.

In addition to rule-action pairs, we need a mechanism for connecting
nonterminals to types in the host programming language. We accomplish
this by simply adding syntax to map a nonterminal to a type. For
example, to abbreviate the Typed Racket \texttt{Integer} type as
\texttt{Int}, one would write the following in a grammar.
\begin{alltt}
  Int = Integer;
\end{alltt}

The implementation of our parsing system translates an input program,
containing a mixture of Typed Racket plus our grammar extensions, into
a program that is purely Typed Racket. In the following we describe
the translation.

A nonterminal-type mapping is translated into a type alias definition.
So $A\ \texttt{=}\ T\texttt{;}$ translates to
\begin{alltt}
  (define-type \(A\) \(T\))
\end{alltt}
where $T$ is an expression that evaluates to a type in Typed Racket.

We use two auxiliary functions to compute the arguments of
rule-functions and rule-macros for translation.
The \emph{support} of a sequence $\alpha$ is the sequence of variables
bound in $\alpha$; the \emph{binders} of $\alpha$ is the sequence of
variable bindings in $\alpha$. In the following definitions
we use list comprehension notation.
\begin{align*}
  \mathit{supp}(\alpha) &= [ x_i \mid \alpha_i \in \alpha, \alpha_i = x_i:B_i ] \\
  \mathit{binders}(\alpha) &= [ x_i:B_i \mid \alpha_i \in \alpha, \alpha_i = x_i:B_i ]
\end{align*}

For both rule-functions and rule-macros, our system generates a unique
name $f$ and $m$, respectively, for use in the Typed Racket output.
Then a rule-function of the form $\forall\overline{x}.\, A^f \to
\alpha => e$ is translated to the definition:
\begin{alltt}
  (: \(f\) (All (\(\overline{x}\)) (\(\overline{B}\) -> \(A\))))
  (define \(f\) (lambda (\(\mathit{supp}(\alpha)\)) \(e\)))
\end{alltt}
A rule-macro of the form $\forall\overline{x}.\, A^m \to \alpha = e$ is translated to the following:
\begin{alltt}
  (define-syntax \(m\)
    (syntax-rules ()
      ((\(m\) \(\overline{x}\ \mathit{supp}(\alpha)\)) \(e\))))
\end{alltt}
The type parameters $\overline{x}$ are passed as arguments to
macros so they can be used in Typed Racket forms. For example, the rule for
{\tt let} expressions in Figure~\ref{fig:let-example} translates to a
typed-let expression in Racket using the parameter {\tt T1}.

Next we show the translation of parse trees to Typed Racket,
written $\llbracket t \rrbracket$.  The key
idea is that we translate a parse tree for a rule-function pair into a
function application, and a parse tree of a rule-macro pair into a macro
application,
\begin{align*}
  \llbracket A^f \to{t}_{\alpha}\rrbracket &=
  \texttt{(}f\ \overline{\llbracket t_{\alpha'}\rrbracket}\texttt{)} \\
  \llbracket\forall\overline{x}.\ A^m \to^{\sigma}{t}_{\alpha}\rrbracket &=
  \texttt{(}m\ \sigma(\overline{x})\
  \overline{\llbracket t_{\alpha'}\rrbracket}\texttt{)}
\end{align*}
where in each case $\alpha' = \mathit{binders}(\alpha)$.
Terminals simply translate as themselves, $\llbracket a \rrbracket = a$.

\subsection{Examples}

Here we present examples in which we add syntax for two DSLs to the
host language Typed Racket.

\subsubsection{Giving ML-like Syntax to Typed Racket}

The module in Figure~\ref{fig:let-example} gives ML-like syntax to several
operators and forms of the Typed Racket language.  The grammar rules for
\texttt{Int} and \texttt{Id} use regular expressions (in Racket syntax) on
the right-hand side of the grammar rule.

\begin{figure}[tbp]
  \centering
\begin{alltt}
{\bf{module}} ML \{
  {\it{// type aliases}}
  Int = Integer;
  Bool = Boolean;

  {\it{// functions}}
  Int ::= x:Int "+" y:Int [left,1] => (+ x y);
  Int ::= x:Int "*" y:Int [left,2] => (* x y);
  Bool ::= x:Int "<" y:Int => (< x y);
  forall T. 
    Void ::= "print" x:T ";" => (displayln x);

  {\it{// macros}}
  forall T.
    T ::= "if" t:Bool "then" e1:T "else" e2:T =
      (if t e1 e2);
  forall T1 T2.
    T2 ::= "let" x:Id "=" y:T1 "\{" x:T1; z:T2 "\}" =
      (let: ([x : T1 y]) z);
  forall T1 T2.
    T2 ::= e1:T1 e2:T2 [left] = 
      (begin e1 e2);

  {\it{// tokens}}
  Int ::= #rx"^[0-9]+$";
  Id ::= #rx"^[a-zA-Z][a-zA-Z0-9]*$";
\}
\end{alltt}
  \caption{An example of giving ML-like syntax to Typed Racket.}
  \label{fig:let-example}
\end{figure}

We then use this module in the following program and save it to the
file {\tt let.es}:
\begin{alltt}
  {\bf{import}} ML;
  let n = 7 \{
    if n < 3 then print 6;
    else print 2 + n * 5 + 5;
  \}
\end{alltt}
Then we compile it and run the generated {\tt let.rkt} by entering
\begin{alltt}
  \$ esc let.es
  \$ racket -I typed/racket -t let.rkt -m
  42
\end{alltt}
where {\tt esc} is our extensible syntax compiler. The result,
of course, is \texttt{42}.

\subsubsection{A DSL for Sets}

The module below defines syntax for converting lists to sets,
computing the union and intersection of two sets, and the cardinality
of a set. Each rule-macro expands to a Racket library call.


\begin{alltt}
  {\bf{module}} Sets \{
    Int = Integer;
    Set = (Setof Int);
    List = (Listof Int);

    Set ::= "\{" xs:List "\}" = (list->set xs);
    List ::= x:Int = (list x);
    List ::= x:Int "," xs:List = (cons x xs);
    
    Set ::= s1:Set "|" s2:Set [left,1] =
      (set-union s1 s2);
    Set ::= s1:Set "\&" s2:Set [left,2] =
      (set-intersect s1 s2);
    Int ::= "|" s:Set "|" = (set-count s);
  \}
\end{alltt}

After importing this DSL, programmers can use the set syntax directly in
Typed Racket. We can also combine the {\tt Sets} module with the {\tt ML}
module from before, for example:
\begin{alltt}
  {\bf{import}} ML, Sets;
  let A = \{1, 2, 3\} \{
    let B = \{2, 3, 4\} \{
      let C = \{3, 4, 5\} \{
        print |A \& C|;
        print A | B \& C;
      \}
    \}
  \}
\end{alltt}
Saving this program in {\tt sets.es}, we can then compile and run it:
\begin{alltt}
  \$ esc sets.es
  \$ racket -I typed/racket -t sets.rkt -m
  1
  \#<set: 1 2 3 4>
\end{alltt}

\section{Related Work}
\label{sec:related-work}

There have been numerous approaches to extensible syntax for
programming languages. In this section, we summarize the approaches
and discuss how they relate to our work. We organize this discussion
in a roughly chronological order.

In the Lithe language, \citet{Sandberg1982} merges the notion of
grammar rule and macro definition and integrates parsing and type
checking. Unfortunately, he does not describe his parsing algorithm.
%
%
\citet{Aasa1988} augments the ML language with extensible syntax for
dealing with algebraic data types. They develop a generalization of
the Earley algorithm that performs Hindley-Milner type inference
during parsing.  However, \citet{Pettersson1992} report that the
algorithm was too inefficient to be practically usable.
\citet{Pettersson1992} build a more efficient system based on LR(1)
parsing. Of course, LR(1) parsing is not suitable for our purposes
because LR(1) is not closed under union, which we need to compose
DSLs.
Several later works also integrate type inference into the Earley
algorithm~\cite{Missura1997,Wieland2009}.
It may be possible to adapt these ideas to enable our approach to
handle languages with type inference.

\citet{Cardelli1994} develop a system with extensible syntax and
lexical scoping. That is, their system supports syntax extensions that
introduce variable binders. Their work inspires our treatment of
variable binders in Section~\ref{sec:binders}.  \citet{Cardelli1994}
base their algorithm on LL(1) parsing, which is also not closed under
union. Also, their system differs from ours in that parsing and
type checking are separate phases.
The OCaml language comes with a preprocessor, Camlp4, that provides
extensible syntax~\cite{Rauglaudre:2002vn}. The parsing algorithm in
Camlp4 is ``something close to LL(1)''. 

\citet{Goguen1992} provide extensible syntax in the OBJ3 language in
the form of mixfix operators. In OBJ3, types (or sorts) play some role
in disambiguation, but their papers do not describe the parsing
algorithm. There is more literature regarding
Maude~\cite{Clavel:1999ys}, one of the descendents of OBJ3.  Maude
uses the SCP algorithm of \citet{Quesada:1998zr}, which is bottom-up
and bidirectional, much like our island parser. However, we have not
been able to find a paper that describes how types are used for
disambiguation in the Maude parser.

The Isabelle Proof Assistant~\cite{Paulson:1994vn} provides support
for mixfix syntax definitions. The algorithm is a variant of chart
parsing and can parse arbitrary CFGs, including ambiguous ones.  When
there is ambiguity, a parse forest is generated and then a later type
checking pass (based on Hindley-Milner type inference) prunes out the
ill-typed trees.

\citet{Ranta2004} develops the Grammatical Framework which integrates
context free grammars with a logical framework based on type theory,
that is, a rich type system with dependent types. His framework
handles grammar rules with variable binders by use of higher-order
abstract syntax. The implementation uses the Earley algorithm and type
checks after parsing, similar to Isabelle.

Several systems use Ford's Parsing Expression Grammar (PEG)
formalism~\cite{Ford2004}. PEGs are stylistically similar to CFGs;
however, PEGs avoid ambiguity by introducing a prioritized choice
operator for rule alternatives and PEGs disallow left-recursive rules.
We claim that these two restrictions are not appropriate for composing
DSLs. The order in which DSLs are imported should not matter and DSL
authors should be free to use left recursion if that is the most
natural way to express their grammar.

\citet{Danielsson:2008kx} investigate support for mixfix operators for
Agda using parser combinators with memoization, which is roughly
equivalent to the Earley algorithm.  Their algorithm does not use
type-based disambiguation during parsing, but they note that a
type-checking post-processor could be used to filter parse trees, as
is done in Isabelle.

The MetaBorg~\cite{Bravenboer2005b} system provides extensible syntax
in support of embedding DSLs in general purpose languages.  MetaBorg
is built on the Stratego/XT toolset which in turn used the syntax
definition framework SDF~\cite{Heering1989} which uses scannerless GLR
to parse arbitrary CFGs. Like Isabelle, the MetaBorg system performs
type-based disambiguation to prune ill-typed parse trees from the
resulting parse forest. Our treatment precedence and associativity is
based on their notion of disambiguation filter~\cite{Brand2002}.
We plan to explore the scannerless approach in the future.
\citet{Bravenboer:2009} look into the problem of composing DSLs and
investigate methods for composing parse tables. We currently do not
create parse tables, but we may use these ideas in the future to
further optimize the efficiency of our algorithm.


\citet{Jim:2010ve} develop a grammar formalism and parsing algorithm
to handle data-dependent grammars. One of the contributions of their
work is ability to bind parsing results to variables that can then be
used to control parsing. We use this idea in Section~\ref{sec:binders}
to enable grammar rules with variable binding. Their algorithm is a
variation of the Earley algorithm and does not perform type-based
disambiguation but it does provide attribute-directed parsing.



\section{Conclusions}
\label{sec:conclusions}

In this paper we presented a new parsing algorithm,
\emph{type-oriented island parsing}, that is the first parsing
algorithm to be constant time with respect to the size of the grammar
under the assumption that the grammar is sparse. (Most parsing
algorithms are linear with respect to the size of the grammar.)  Our
motivation for developing this algorithm comes from the desire to
compose domain-specific languages, that is, to simultaneously
import many DSLs into a software application.

We present an extensible parsing system that provides a front-end
to a host language, such as Typed Racket, enabling the definition of
macros and functions together with grammar rules that provide
syntactic sugar. Our parsing system provides precedence and
associativity annotations, parameterized grammar rules, and grammar
rules with variable binders and scope.

In the future we plan to extend the syntax of nonterminals to
represent structural types, formally prove the correctness of
type-oriented island parsing, pursue further opportunities to improve
the performance of the algorithm, and provide diagnostics for helping
programmers resolve the remaining ambiguities that are not addressed
by typed-based disambiguation.

\bibliographystyle{abbrvnat}
\bibliography{refs}
\end{document}